\documentstyle[psfig,epsf,rotate]{mn}

\title[VLT spectroscopy of XTE J2123-058: the masses]
{VLT spectroscopy of XTE J2123-058 during quiescence: the masses of the two 
components}

\author[J. Casares et al.]{
J. Casares$^1$, G. Dubus$^2$, T. Shahbaz$^1$, C. Zurita$^1$, 
P.A. Charles$^3$\\
$^1$Instituto de Astrof\'\i{}sica de Canarias, 38200 La Laguna, Tenerife, 
Spain\\
$^2$Theoretical Astrophysics, Caltech 130-33, CA 91125, Pasadena\\
$^3$Dept of Physics \& Astronomy, University of Southampton, Southampton, UK}
\begin{document}

\maketitle

\begin{abstract}

We present VLT low resolution spectroscopy of the neutron star X-ray
transient XTE J2123-058 during its quiescent state. Our data
reveal the presence of a K7V companion which contributes 77 \% to the
total flux at $\lambda$6300 and orbits the neutron star at $K_2 =
287 \pm 12$ km s$^{-1}$. Contrary to other soft X-ray transients (SXTs), 
the $H_{\alpha}$ emission is almost exactly in antiphase with the velocity 
curve of the
optical companion. Using the light-center technique we obtain $K_1 =
140 \pm 27$ km s$^{-1}$ and hence $q=K_1/K_2=M_2/M_1= 0.49 \pm
0.10$. This, combined with a previous determination of the inclination
angle ($i=73^{\circ} \pm 4$) yields $M_1 = 1.55 \pm 0.31$ M$_{\odot}$
and $M_2 = 0.76 \pm 0.22$ M$_{\odot}$.  $M_2$ agrees well with the
observed spectral type. Doppler tomography of the H$_{\alpha}$
emission shows a non-symmetric accretion disc distribution mimicking 
that seen in SW Sex stars. Although we find a large systemic velocity of 
-110 $\pm 8$ km s$^{-1}$ this value is consistent with the galactic rotation 
velocity at the position of J2123-058, and hence a halo origin. The formation 
scenario of J2123-058 is still unresolved.
  
\end{abstract}


\section{INTRODUCTION}

Mass determinations of neutron stars (NS) can provide essential
constraints on the equation of state of nuclear matter (van Kerkwijk,
van Paradijs \& Zuiderwijk 1995). Compact object masses are now 
known accurately for radio pulsars, all of which are consistent with
$1.38 \pm$ 0.07 M$_{\odot}$ (Thorsett et al. 1993), but have been very
difficult to obtain in low mass X-ray binaries (LMXBs) both because
their NS do not pulse and because the optical flux from the companion
is usually overwhelmed by reprocessed X-ray emission from the accretion disk
(van Paradijs \& McClintock 1995). Only in the quiescent state of soft
X-ray transients (SXTs) can the faint companion be detected and
dynamical information extracted. 
An accurate constraint on the NS mass, however, requires knowledge of 
both the system inclination $i$ and the mass ratio $q$ 
in addition to the amplitude of the radial
velocity curve of the companion $K_2$.\\

The transient XTE J2123-058, discovered in June 1998, offers an
excellent opportunity to achieve this. The presence of a NS is
confirmed by type-I bursts (Tomsick et al. 1999; Homan et al. 1999,
Gneiding, Steiner \& Cieslinski 1999) which makes it one of only 4 NS
SXTs known. The optical counterpart showed clear triangular-shaped minima
repeating on the orbital period of 6~hrs which consequently evolved to an 
ellipsoidal modulation before the system entered quiescence at
$R=21.7$ (Tomsick et al. 1999; Soria, Wu \& Galloway 1999; Zurita et
al. 2000; Hynes et al. 2001). The modulation yielded
$i=73\pm4^{\circ}$ (Zurita et al. 2000) but it should be noted that 
systematic effects related to the assumed outburst irradiation model 
might have some importance. \\

The secondary star is suspected to be of late K-type in order to
transfer matter within the constraints of a 6 hr orbit. Photometric 
colours obtained during 
quiescence also suggest the presence of a late-K companion (see Zurita
et al 2000, Shahbaz et al. 2001). In this paper we report the first
spectroscopic detection of the companion star in J2123-058 and its 
radial velocity curve using the {\em Kueyen} VLT (sections \S3 and \S4). 
Sections \S5 and \S6 deal with the analysis of the $H_{\alpha}$ 
emission. In section \S7 we present the spectroscopic constraints on 
the components masses, together with implications for the disc 
instability model and discuss constraints on the formation scenario 
implied by the observed systemic velocity.

\section{OBSERVATIONS AND DATA REDUCTION}

We obtained a total of 20 spectra of J2123-058 during observations
with the FORS-2 spectrometer of the {\em Kueyen} VLT on August 23 and
24, 2000 (UT).  We used grism 600R (with the order-sorting filter
GG435+31), centered at 6300~\AA\ with a 0.7" slit on the first night. On the
second night, the seeing deteriorated (up to 1.2") and so we decided to use
a 1 arcsec slit. The spectral
resolution, as measured from the FWHM of the arclines, was 2.7~\AA~ and
4.0~\AA~ for the two nights, respectively. Exposure times of 2200 s
were employed.\\

The data were de-biased and flat-fielded in the usual way. The spectra
for J2123-058 were then extracted using standard optimal extraction
techniques which optimize the signal-to-noise ratio of the final
spectra (Horne 1986).  Calibration HeNeAr arc images were obtained in
daytime with the telescope parked at the zenith. The
$\lambda$-pixel scale was derived through fourth-order polynomial fits
to 40 lines resulting in an rms scatter better than 0.3 \AA. Internal
flexure was measured by cross-correlation of the sky spectra and 
was found to be $<$ 20 km s$^{-1}$ ($<$ 0.4 pixel) throughout the
night. Relative shifts of every target spectrum to the first one were
applied and the zero point (-18 $\pm$ 2 km s$^{-1}$) obtained
through a gaussian fit to the strong $\lambda$6300.3 OI line on the
first sky spectrum. The spectra were subsequently rebinned into a
constant velocity scale of 52 km s$^{-1}$ pixel$^{-1}$. \\

Spectra of several K and M-type giants (table 1) were obtained with the same
set-up (slit width=0.7") for the cross-correlation analysis. A
database of K-type main sequence stars (table 2), observed with the IDS on 
the INT with similar wavelength coverage and resolution (see Casares et
al. 1996) were also used for the spectral type classification.

\section{RADIAL VELOCITY CURVES AND EPHEMERIS}

Prior to the cross-correlation the spectra were rectified to the
continuum (which was subtracted afterwards) by masking the emission
lines (H$\alpha$, HeI $\lambda$5875, $\lambda$6678) and fitting a
low-order spline to the remainder. Then, individual radial velocities
were extracted by cross-correlation with each template star in the
range $\lambda\lambda$5600-6520, after masking out the atmospheric and
interstellar band at $\lambda\lambda$6270-6320. Note that the NaD feature 
at $\lambda\lambda$5890-96 was included in the cross-correlation because is 
a strong feature in late-type stars and, given the high galactic latitude of 
J2123-058, this is only marginally affected by interstellar absorption; the EW 
of the interstellar NaD measured by Hynes et al. (2001) is less than 15 \% of 
our observed value of 4.2 \AA~ (see Fig. 2). Its effect is less than 1 
km$^{-1}$ on the radial velocity curve. A few spectra from the
first night have very poor statistics and consequently yield unreliable radial
velocities.  This was improved by co-adding the individual
spectra into 16 phase bins, using the established orbital period of 
$P= 0.24821$ days
(Zurita {\it et al.} 2000). The new radial velocities were fitted with 
a sine-wave at this period, the resulting parameters being 
displayed in Table 1.

\begin{table*}
\caption{Radial Velocity Parameters for J2123-058}
\begin{center}
\vspace{10pt}
\begin{tabular}{lccccc}
 
{\em Template}&{\em Spectral Type}&{\em $\gamma_{2}^{\dagger}$} & 
{\em $T_{0}^{\dagger\dagger}$} & {\em $K_{\rm 2}$} & {\em $\chi^{2}_{\nu}$} \\
{\em Star} & & {\em (km s$^{-1}$)} & {\em (+ 2451779.0)} & {\em (km s$^{-1}$)} & \\
\\
HR 5265 & K3 III &-112$\pm$10&0.6515 $\pm$ 0.0015 & 285 $\pm$ 9 & 1.8 \\
HR 5178 & K5 III &-109$\pm$11&0.6520 $\pm$ 0.0012 & 289 $\pm$ 9 & 1.8 \\
HR 6159 & K7 III &-111$\pm$11&0.6520 $\pm$ 0.0017 & 287 $\pm$ 12 & 1.3 \\
HR 5496 & M1 III &-108$\pm$13&0.6515 $\pm$ 0.0015 & 290 $\pm$ 10 & 1.7 \\
HR 5603 & M3 III &-111$\pm$15&0.6517 $\pm$ 0.0017 & 287 $\pm$ 11 & 1.6 \\
\end{tabular}
\vspace{5pt}
\newline
\noindent
{\footnotesize{$\dagger$ Instrumental flexure and radial velocities of 
templates have been removed.}}\\
\noindent
{\footnotesize{$\dagger\dagger$ This refers to the inferior conjunction of the 
secondary star.}}
\end{center}
\end{table*}

Column 3 lists the systemic velocities after correcting for the
template radial velocities (as given by SIMBAD) and instrumental flexure 
on template spectra. This latter correction could not be
accounted for in the same way as in the target spectra because of the
faintness of the sky spectra. Instead, this was estimated by simply 
measuring the central wavelength of the H$_{\alpha}$ line, by fitting 
gaussians to the line core. 
Notice the small scatter in the distribution of $\gamma$-velocities 
which render confidence in the technique applied to compensate for 
differential flexure corrections. The statistical mean 
is -110 $\pm$ 1 km s$^{-1}$ but the real
error is probably underestimated. As a better representation of the
uncertainty in the true systemic velocity we prefer to quote the mean
of 1-$\sigma$ uncertainties of the fits to the radial velocity curves. 
Therefore, our favoured determination of the systemic
velocity in J2123-058 is $\gamma_2 = -110 \pm 8$ km s$^{-1}$. \\

Note the consistent values of the other parameters in Table 1 which
suggest $T_0 = HJD 2451779.652 \pm 0.001$ and $K_{2} = 285-290$ km
s$^{-1}$ (all quoted errors are $\pm 1\sigma$).  The combination of our
$T_0$ determination with that of Zurita et al. (2000) enables us to refine the
orbital period to $P=0.248236 \pm 2 \times 10^{-6}$ d. We note that
HR6159 yields a significantly lower $\chi^2_{\nu}$ than the rest and,
therefore, we adopt the orbital parameters derived using this
template.  Adopting $K_2 = 287 \pm 12$ km s$^{-1}$, the implied mass
function is $f(M) = 0.61 \pm 0.08$ M$_{\odot}$. The phase-folded radial
velocity curve, for the case of the K7 III template, is shown in the
top panel of Fig. 1.

\section{SPECTRAL CLASSIFICATION AND ROTATIONAL BROADENING}

Figure 2 presents the Doppler-corrected average spectrum of J2123-058
in the rest frame of the secondary, together with some spectral-type
standards. Different weights were assigned to different spectra in
order to maximize the signal-to-noise ratio. Photospheric absorption 
features from the companion are clearly visible, such
as the metallic blends at $\lambda$6165 and $\lambda$6495 and the NaI
doublet at $\lambda$5890-6. These are clearly broader in J2123-058 than 
in the luminosity class III templates but this is purely an 
instrumental effect since the averaged spectrum is dominated by the
higher quality spectra from the second night, obtained through a 1"
slit. Several molecular TiO bands (e.g. $\lambda\lambda$6150-6250,
$\lambda\lambda$6960-90 and $\lambda\lambda$7020-50) are also evident
and these can be used to rule out spectral types K3 (and earlier) and
M1 (and later). The spectral type of the companion is most probably in
the range K5-7. \\

Both the spectral type and luminosity class of the secondary can be
efficiently determined by subtracting different broadened versions of
the templates (to compensate for the instrumental resolution mismatch)
from the Doppler-corrected sum of the target until the lowest residual
is obtained (as performed in e.g. Marsh, Robinson \& Wood 1994). The 
broadened templates are multiplied by a factor $f\leq 1$ to account for 
the shallowness of the absorption features in J2123-058, which are diluted 
by the continuum excess of the accretion disc. The results of the optimal
subtraction analysis are presented in Table 2.  As expected, main
sequence templates yield systematically lower $\chi^2$, with a minimum
for spectral types K7-8 V.  The relative contribution of the
companion star to the total flux in the R-band is constrained to the
range $f=0.70-0.77$.

\begin{table*}
\caption{Spectral Classification}
\begin{center}
\vspace{10pt}
\begin{tabular}{lcccc}
{\it Template} & {\em Spectral} & {\it Broadening} & {\em $f$} & {\em
$\chi^2_{\nu}$}\\ 
& {\em Type} & {\em (km s$^{-1}$)} & & (d.o.f.=632)   \\
\\
HR 5265 & K3 III & 230  & 1.00 $\pm$  0.14 &  0.300  \\
HR 5178 & K5 III & 210  & 0.85 $\pm$  0.11 &  0.293  \\
HR 6159 & K7 III & 200  & 0.84 $\pm$  0.11 &  0.300  \\
HR 5496 & M1 III & 155  & 0.68 $\pm$  0.10 &  0.312  \\
HR 5603 & M3 III & 120  & 0.52 $\pm$  0.08 &  0.324  \\
\\
HD 184467   &  K2 V & 190 & 1.43 $\pm$ 0.17 &  0.267 \\
HD 154712 A &  K4 V & 170 & 0.90 $\pm$ 0.10 &  0.263 \\
HD 296974   &  K5 V & 185 & 0.83 $\pm$ 0.09 &  0.257 \\
 61 Cyg A   &  K5 V & 220 & 0.83 $\pm$ 0.09 &  0.258 \\
 61 Cyg B   &  K7 V & 185 & 0.77 $\pm$ 0.08 &  0.255 \\
HD 154712 B &  K8 V & 190 & 0.70 $\pm$ 0.07 &  0.255 \\
\end{tabular}
\end{center}
\end{table*}

A rough estimate of the rotation speed of the companion star (assuming
co-rotation) is provided by the size of the Roche lobe ($R_{\rm L2}$),
which is related to the secondary's mass ($M_{\rm 2}$) and the orbital
period ($P$) by combining Kepler's Third Law with Paczynski's
expression (Paczynski 1971) for the equivalent radius of a Roche lobe
filling star.

$$
\left(\frac{R_{\rm L2}}{\rm R_{\odot}}\right) = 0.234 
\left(\frac{P}{\rm hr}\right)^{2/3} \left(\frac{M_{\rm 2}}{M_{\odot}}
\right)^{1/3}
$$

\noindent
An upper limit will be obtained by adopting the nominal mass of a K7 V
main sequence star (0.65 $M_{\odot}$) because the companions in close
binaries are usually undermassive for their spectral types (van den Heuvel
1983). With $R_{\rm L2} \leq 0.67 R_{\odot}$, the secondary would just
fill its Roche lobe or be slightly evolved (a K7V has $R_{2} \simeq
0.69 R_{\odot}$). This implies a maximum rotation speed of $\simeq
139$ km s$^{-1}$, but the projected rotational broadening ($V_{\rm
rot} \sin i$) cannot be much less than this because of the high
inclination ($i\simeq73^{\circ}$, see Zurita et al. 2000). Since this
is smaller than the spectral resolution on 23 Aug (185 km s$^{-1}$),
and because of the orbital smearing due to our long integration times
($\simeq 0.1 P$), we cannot measure $V_{\rm rot} \sin i$ in our
Doppler-corrected sum. However, using the limit that $V_{\rm rot} <
139$ km s$^{-1}$ and $i\simeq73^{\circ}$ then the
mass ratio is constrained to be $q=M_{\rm 2}/M_{\rm 1} \le 0.47$ (Wade
\& Horne 1988).

\section{THE MASS RATIO}

The system mass ratio can also be determined if the radial velocity
curve of the compact object is available. Unfortunately, the neutron
star in J2123-058 does not pulse and we can therefore only hope for the
velocities of the extreme H$_{\alpha}$ wings (which originate in the
vicinity of the compact star) to share its orbital motion. The
``double-gaussian'' technique allows us to extract the centroid of
line velocity regions $\pm v$ through convolving the individual
spectra with two Gaussian bandpasses separated by $a=2v$ (Schneider \&
Young 1980). For a given value of {\it a} the velocity points can be
fitted by a velocity curve of the form

$$V= \gamma + K \sin 2 \pi (\phi - \phi_0 ) = 
\gamma - K_x \cos 2 \pi \phi + K_y \sin 2 \pi \phi$$

\noindent
and ($K_x$,$K_y$) be represented in velocity space. If the inner disc were 
axisymmetric $K_x = 0$, $K_y=-K_1$ and one could recover the radial velocity
curve of the compact star. The application of this technique to
cataclysmic variables (CVs) and black-hole X-ray transients (BHXRTs) 
has shown that emission line radial velocity curves are systematically 
delayed (i.e. $K_x < 0$) and also their $\gamma$-velocities differ from 
the systemic velocity (probably due to mass outflows) casting doubts on any 
dynamical interpretation (Marsh 1998). To overcome this problem, Marsh 
(1988) proposed the {\it lightcenter} technique which basically extrapolates 
the ($K_x$,$K_y$) points towards the vertical axis in velocity space 
(i.e. to absolute phase 0) before values get corrupted by noise at large 
gaussian separation. 

\begin{table*}
\caption{Double-Gaussian analysis of the H$_{\alpha}$ emission line}
\begin{center}
\vspace{10pt}
\begin{tabular}{lcccccc}
{\it Separation $a$} & {\em $\chi_{\nu}^2$} & {\it $\gamma$} & {\em $\phi_0$} & 
{\em $K_x$} & {\em $K_y$} &  {\em $K=\sqrt{K_x^2 + K_y^2}$} \\
 {\em (km s$^{-1}$)} &  & {\em (km s$^{-1}$)} & & {\em (km s$^{-1}$)} & 
 {\em (km s$^{-1}$)} &  {\em (km s$^{-1}$)} \\
\\
 400 & 3.0 & -106$\pm$19 & 0.532$\pm$0.010 & -90$\pm$33 &-442$\pm$49 & 451$\pm$25\\
 600 & 1.0 & -105$\pm$26 & 0.548$\pm$0.013 &-122$\pm$42 &-393$\pm$67 & 412$\pm$33\\
 800 & 2.8 & -144$\pm$17 & 0.508$\pm$0.017 & -14$\pm$30 &-269$\pm$41 & 269$\pm$22\\
1000 & 1.4 & -123$\pm$18 & 0.487$\pm$0.022 &  15$\pm$27 &-178$\pm$44 & 179$\pm$24\\
1200 & 0.6 & -120$\pm$13 & 0.500$\pm$0.017 &   0$\pm$17 &-155$\pm$33 & 155$\pm$20\\
1400 & 0.3 & -109$\pm$17 & 0.523$\pm$0.022 & -20$\pm$23 &-139$\pm$49 & 140$\pm$29\\
1600 & 0.2 & -145$\pm$20 & 0.592$\pm$0.034 & -76$\pm$40 &-117$\pm$59 & 140$\pm$27\\
1800 & 0.5 & -124$\pm$25 & 0.674$\pm$0.036 &-132$\pm$43 & -68$\pm$69 & 149$\pm$31\\
2000 & 4.1 & -184$\pm$26 & 0.582$\pm$0.022 &-113$\pm$46 &-199$\pm$75 & 229$\pm$38\\
\end{tabular}
\end{center}
\end{table*}

We have followed this approach and 
applied the "double-gaussian" technique to our 16 phase-binned
spectra of J2123-058 using bandpass separations from 400 to 2000 km
s$^{-1}$ and $\sigma=200$ km s$^{-1}$ of gaussians to match our
instrumental resolution. The results of the fits are listed in table 3
and overplotted on the Doppler map (see next Sect.) in Fig. 3. As we
move away from the line core, the radial velocity amplitude $K
(=\sqrt{K_{x}^2 + K_{y}^2})$ decreases from 450 to $\sim 140$ km
s$^{-1}$ and the phase lag, relative to the expected motion of the
compact star, quickly vanishes (i.e. $\phi_0 \rightarrow 0.5$). Further into the 
wings the points are
again dragged towards large (negative) velocities as if they were
contaminated by a high velocity component. This behaviour is different
from other BHXRTs (e.g. Marsh, Robinson \& Wood
1994) where contamination by the hot-spot distorts the emission line
velocities all the way into the extreme wings. At this point we
tentatively adopt $K_1 = 140 \pm 29$ km s$^{-1}$ as the lowest
$K$-amplitude that is consistent with the correct phasing of the compact
object ($\phi_0 = 0.5$). This is fulfilled for a gaussian separation 
$a\simeq1400$ km s$^{-1}$, which also yields a $\gamma$-velocity in good 
agreement with the companion star (see Sect 3). This radial velocity curve 
for H$\alpha$ is shown in the middle panel of Fig. 1.

\section{DOPPLER IMAGE OF THE H$\alpha$ EMISSION}   

The mean spectrum of J2123-058 in the laboratory rest frame is dominated by 
broad (FWHM=1300 km s$^{-1}$), strong (EW=20 \AA), double-peaked H$\alpha$ 
emission. The EW traces a $\simeq$ 20 \% amplitude
sinusoidal modulation with maxima at phase 0 i.e. inferior conjunction
of the neutron star (see lower panel of Fig. 1). \\
 
We have combined our phase-binned H$\alpha$ profiles to reconstruct
the brightness distribution of the accretion disc and other emission
sources in the system using the Doppler Tomography technique (Marsh \&
Horne 1988). The continua were normalized and subtracted so as to give
a pure emission line profile which was subsequently rebinned onto a
constant velocity scale (52 km s$^{-1}$ pix$^{-1}$) covering 58 pixels
centered at H$\alpha$. The resulting doppler map is presented in Fig.4
in velocity coordinates, together with the trailed spectra. The map is
dominated by a crescent-shaped structure, pointing at phase $\simeq$
0.5, superposed on the weak ring-like structure of the accretion
disc with no sign of emission at the expected
hot-spot location (the theoretical path of the gas stream for $q=0.49$
is marked in units of $R_{\rm L1}$). 

We do see some emission from the $L_1$ point. Similar emission 
has been reported in other quiescent X-ray transients, N. Mus 91
(Casares et al. 1997), N. Oph 77 (Harlaftis et al. 1997), Cen X-4
(Torres et al. 2001) which may be indicative of chromospheric
activity in these late-K companions. However, the highly asymmetric
brightness distribution of the accretion disc emission in J2123-058 is 
markedly distinct from other BHXRTs. The predominance of emissivity
from the lower quadrants in the Doppler map is more reminiscent of
CVs of the SW Sex type, which may suggest that a similar
accretion processes (magnetic accretion, disc overflow or propeller)
is at work.  This similarity was already noted by Hynes et
al. (2000) using better quality spectroscopy of J2123-058 during the
outburst phase.

\section{DISCUSSION}

J2123-058 is a key system for the study of the physics of neutron stars 
and X-ray transients. It is one of only a handful of neutron star transients 
and is at high inclination. This latter property will eventually lead to an  
accurate measurent the masses of both stars. 
This work presents the first spectroscopic analysis of J2123-058 in
quiescence, with a determination of the companion's radial velocity
curve.  Contrary to other SXTs, the H$\alpha$ emission is antiphased
with the companion star. Consequently, we have employed the
Double-gaussian technique to estimate $K_1$ and hence the binary mass
ratio $q = K_1/K_2=M_{2}/M_{1}$.  In Table 4 we list our determination
of the system parameters in J2123-058.  From the mass function and $q$
we constrain the compact object mass to be $M_{1} \sin^{3} i = 1.35 \pm
0.25$ M$_{\odot}$. The inclination was determined in Zurita et
al. (2000) by fitting the outburst lightcurves with an X-ray heated
model. Using $i=73 \pm 4^{\circ}$ in our mass function we get
$M_{1}=1.55 \pm 0.31$ M$_{\odot}$ and $M_{2}=0.76 \pm 0.22$
M$_{\odot}$, which can be further constrained to $M_{2}=0.54-0.8$ 
using the strong spectroscopic evidence that the companion cannot 
possibly be a K3V or earlier.

\begin{table*}
\caption{Orbital Parameters} 
\begin{center}
\vspace{10pt}
\begin{tabular}{ll}
{\it Parameter} & {\it Values} \\
{\em Period (days)} & 0.248236 $\pm$ 0.000002 \\
{\em K$_{2}$ (km s$^{-1}$)} & 287 $\pm$ 12 \\
{\em $\gamma$ (km s$^{-1}$)} & -110 $\pm$ 8 \\
{\em T$_{\circ}$ (HJD)} & 2451779.652 $\pm$ 0.001 \\
{\em q (=K$_x$/K$_c$)} & 0.49 $\pm$ 0.10  \\
{\em f(M$_x$/M$_{\odot}$)} & 0.61 $\pm$ 0.08 \\
{\em i } & 73 $\pm 4^{\circ}$ \\
{\em M$_{1}$ (M$_{\odot}$)} & 1.55 $\pm$ 0.31 \\
{\em M$_{2}$ (M$_{\odot}$)} & 0.76 $\pm$ 0.22 \\
\end{tabular}
\end{center}
\end{table*}

The compact object mass is consistent with the canonical value of
standard neutron stars although our $1-\sigma$ uncertainty allows as
large a mass as $\simeq$ 1.9 M$_{\odot}$. To set constraints on the
equation of state of nuclear matter will require refining
$f(M)$ and, in particular, $q$ and $i$.  A 50\% improvement in the
errorbars of $q$ and $i$ will therefore provide a significant gain and 
this can be attempted through detailed modeling of the optical lightcurves 
both in outburst and quiescence.\\

The spectrum and mass of the companion star are both 
consistent with a "normal" K7 main sequence star. With an orbital period 
of 6 hours, the mass transfer rate from the K7V star is expected to be set 
by orbital angular momentum losses due to magnetic braking.  For a Roche 
lobe filling main sequence star this leads to $\dot{M}_2\approx 2\cdot
10^{-9}$ M$_\odot$~yr$^{-1}$ (Eq. 16 in King 1988, see also Fig. 1 of
King, Kolb \& Burderi 1996). 
On the other hand, the critical mass accretion rate above which the
disk becomes stable against the thermal-viscous instability is
$\dot{M}_{\rm crit}^{\rm irr}\approx 3\cdot10^{-10}$
M$_\odot$~yr$^{-1}$ if irradiation heating is included and
$\dot{M}_{\rm crit}^{\rm no~irr}\approx 3\cdot10^{-9}$
M$_\odot$~yr$^{-1}$ when it is not (Fig. 10, Dubus et al. 1999). The
critical mass accretion rate is independent of the assumed magnitude
of the viscosity (Eq. 39 of Hameury et al. 1998). In the most
favourable case with negligible irradiation heating (unlikely
considering the interpretation of the orbital modulation observed in
outburst that required an irradiated disk), the accretion disk is only
marginally unstable.  The estimated mass transfer rate is close to
$\dot{M}_{\rm crit}$ in which case models (Dubus, Hameury \& Lasota
2001) predict long outbursts and short quiescent periods, contrary to the
behaviour of XTE J2123-058. Magnetic braking thus leads to uncomfortably high
estimates for the mass transfer rate. This problem is common to short
period ($P_{\rm orb}<1-2$~days) neutron star X-ray binaries and has
led King, Kolb \& Burderi (1996) to propose that their companions are
highly evolved and not main-sequence stars.\\

This can be ruled out for XTE J2123-058: $\dot{M}_2<\dot{M}_{\rm
crit}^{\rm irr}$ implies $\hat{m}_2 m_1^{-2/7}< 0.27$ with
$m_1=M_1/$M$_\odot$ (Eq. 9 of King, Kolb \& Burderi 1996 with 
gravitational radiation neglected) and $\hat{m}_2=M_2/M_2(MS)$
where $M_2(MS)$ is the mass of a main-sequence star with the same mean
density as the companion (Kolb, King \& Baraffe 2001); with
$M_2(MS)=0.60$ M$_\odot$ and $m_1=1.55$ this gives $M_2<0.18$
M$_\odot$ which disagrees with our spectroscopic mass of $M_2=0.54-0.98$
M$_\odot$ and spectral type determination. If $M_2=0.18$ M$_\odot$,
the companion might be a stripped giant (see Taam 1983; Webbink, Rappaport \& 
Savonije 1983) consisting of
a degenerate He core and an extended H shell-burning envelope, 
but the implied temperature would then be $T_{\rm eff}\approx 5500$ K rather 
than the observed $T_{\rm eff}\approx 4250$ K (i.e. K7V). Our spectroscopy 
clearly shows that the secondary star cannot possibly be a stripped-giant 
and must be a main sequence K7 or slightly evolved\footnote{Furthermore, 
stripped giants are not possible for periods less than $\sim$ 1 d because 
mass transfer starts before the core mass exceeds the Schonberg-Chandrasekhar 
mass and collapses (King 1988)}.\\

On the other hand, our estimate of $\dot{M}_2$ above from magnetic braking 
implies a {\em mean} luminosity $L\approx \epsilon \dot{M}_2 c^2 \approx 
10^{37}$ ergs$^{-1}$ (with $\epsilon=0.1$) which is comparable to the observed
outburst {\em peak} X-ray luminosity ($L_{\rm x}(2-12{\rm ~keV})\approx 
1.9(\pm0.5)\cdot 10^{37}$ erg s$^{-1}$ for D=9.6 kpc, see below). 
The problem is clearly in how $\dot{M}_2$ was estimated rather than 
in the outburst model. If the secular evolution is driven by gravitational 
radiation losses only, the inferred accretion rate is lower and more 
consistent with the disc instability model ($\dot{M}_2 \approx 5\cdot 
10^{-10}$ M$_\odot$~yr$^{-1}$). Even taking into account angular momentum 
carried away by mass ejected from the system, this cannot significantly 
reduce the estimated $\dot{M_2}$, unless matter is carried away by a highly
relativistic jet close to the NS (King \& Kolb 1999) for which there
is no evidence. \\

The physical mechanisms involved in magnetic braking are poorly known
and it is a distinct possibility that for some reason they do not work
or are less efficient in this system. In fact, even after accounting
for evolved companions, magnetic braking has been shown to give
recurrence times which are much too short for BH SXTs with $P_{\rm
orb}<0.4$ d (Romani 1998). Other assumptions made in the derivation of
$\dot{M}_2$ could be wrong e.g. the secondary may not fill its Roche
lobe or is not corotating, although both are unlikely. Alternatively,
the transient character of XTE J2123-058 could be temporary, for
instance if the donor star undergoes cycles which reduce $\dot{M}_2$
(by at least an order of magnitude) from its secular value.  Such
variations have been invoked to explain the dispersion in mass
accretion rates observed in CVs (Warner 1995).
Possibilities include star spots (Livio \& Pringle 1994) or
instabilities in the secular mass transfer induced by the anisotropic
irradiation of the secondary (Ritter, Zhang \& Kolb 2000).\\

With $R=21.8$ in quiescence, the dereddened apparent magnitude of the
K7V companion is $R=21.9\pm0.3$ (assuming a 30\% $\pm10$\%
veiling). The absolute magnitude of a K7V star is $M_R\approx7$, hence
$D_{\rm kpc}=9.6\pm1.3$, in good agreement with previous estimates (Tomsick 
et al. 1999, Homan et al. 1999). With a distance of $9.6\pm1.3$~kpc and 
$b=-36.2^{\rm o}$ , XTE J2123-058 lies $>5$ kpc above 
the galactic plane and it has been proposed that it may be a halo object 
escaped from a globular cluster (Homan et al. 1999). We can test this 
hypothesis using the new information provided by our systemic radial 
velocity $\gamma_2=-110 \pm 8$ km s$^{-1}$. In the absence of peculiar 
motions, the observed line-of-sight velocity $v_o$ will 
be driven by the differential galactic rotation and the Sun's motion 
relative to the local standard of rest i.e. 
  
$$
v_o = v_{\rm rot} {{R_{\odot} \cos \lambda}\over u} - v_{\odot} \cos \lambda
$$

\noindent
where $u^2=R_{\odot}^2 + D^2 \cos^2 b - 2 D R_{\odot} \cos b \cos l$, 
$\cos \lambda = \cos b \sin l$, $R_{\odot}=8.8$ kpc, $v_{\odot} = 
220$ km s$^{-1}$ and $v_{\rm rot}$ is the galactic rotation velocity  
at the position of J2123-058 (Frenk \& White 1980; Morrison et al. 1990). 
Using $v_o = - 110$ km s$^{-1}$, $b=-36.2^{\rm o}$, $l=46.5^{\rm o}$ and 
$D= 9.6 $~kpc we find $v_{\rm rot} = 24$ km s$^{-1}$ which is consistent 
with the low rotation velocities of globular clusters with galactocentric
distances in the range 2-33 kpc (Frenk \& White 1980) and halo stars 
(Freeman 1988). We note, however, 
that only one velocity component is known and the true rotational velocity 
might be substantially different. In an attempt to find evidence of 
proper motion in J2123-058 we have also performed astrometry on our best 
quality images taken during outburst (5 July 98) and quiescence (19 Aug 99) but 
found no significant motion with upper limits $\mu_{\alpha} < 90~
\rm{mas~yr}^{-1}$ and $\mu_{\delta} < 370~\rm{mas~yr}^{-1}$. Therefore, the 
formation scenario of J2123-058 cannot be tested on pure kinematical grounds. \\

The transition from galactic disc to halo is known to occur sharply at  
[Fe/H] = -1.0 (e.g. Ratnatunga \& Freeman 1989) and hence the companion star 
of J2123-058 would be metal poor ([Fe/H] $< $ -1.0) if formed in the halo. 
Unfortunately, a difference of one dex in metallicity would produce only tiny 
changes in line strength at the low $T_{\rm eff}$ of the secondary. 
These changes are completely undetectable in our moderate S/N average spectrum.
The evidence that J2123-058 is not a halo object was raised by Hynes et al. 
(2001) based on the relative strength of the Bowen blend and HeII 
$\lambda$4686 emission lines during outburst. This supports the idea that 
J2123-058 was indeed formed in the galactic plane and received a large 
supernova kick (150-400 km~s$^{-1}$, Homan et al. 1999) to reach its present
location. We note
that Kalogera, Kolb \& King (1998) conclude that "large kick velocities 
will favour the formation of short-period systems with unevolved donors''. 
\\

Our H$\alpha$ Doppler imaging of the accretion disk poses several
interpretative difficulties. The map shows significant emission at the
$L_1$ point but not from the hot spot. Enhanced emission would be
expected from this hemisphere of the companion during outburst when it
faces irradiation from the accreting neutron star. Typical X-ray luminosities 
of neutron star SXTs in quiescence are $L_{\rm x} \simeq 10^{33}$ erg s$^{-1}$ 
and this luminosity would be sufficient to power the observed H$\alpha$ emission
(more details will be presented in Shahbaz et al. 2001). 
Alternatively, the emission could be due to 
chromospheric activity on the secondary (Bildsten \& Rutledge 2000), e.g. 
star spots close to the $L_1$ point (which would help reduce 
the mass transfer rate) could explain enhanced H$\alpha$ emission. \\

Part of the H$\alpha$ emission observed in Fig.3 is located at velocities 
which are too low to be associated with a Keplerian disk  
and in anti-phase with the 
secondary, a similar situation to that observed during outburst by 
Hynes et al. (2000). Their higher quality Doppler map was very reminiscent 
of those seen in a group of nova-like CVs, the SW 
Sex stars. Hynes et al. (2000) argued that the mechanism was probably 
the same and likely to be a propeller. In this scenario, a magnetic 
field anchored in the disk interacts with the accretion stream from 
the secondary, ejecting blobs of matter out of the system. The 
trajectories depend on the mass but all intersect at some 
location. The collision of the blobs of different sizes in this area 
produces the asymmetric enhanced emission which is observed (Horne 
1999).\\

The SW Sex stars and SXTs in outbursts have high accretion rates. In
contrast, our Doppler map was obtained in quiescence when the
accretion rate is very low. If the outbursts are due to the
thermal-viscous instability, the thin disk becomes neutral which casts
doubts on its ability to anchor a magnetic field for this propeller
mechanism to work. On the other hand, the inner regions of the thin
disk are most probably replaced by a hot thick and optically thin accretion 
flow during quiescence. This is required to explain e.g. the X-ray 
luminosities of quiescent BH SXTs (Lasota, Narayan \& Yi 1996 and the review 
by Narayan, Mahadevan \& Quataert 1998) or the recurrence times of SXTs
(Dubus, Hameury \& Lasota 2001). These ionised flows have very short
accretion timescales which makes it unlikely that they could sustain
large-scale steady magnetic fields reaching to the accretion stream.\\

The neutron star magnetosphere itself can become an efficient
propeller in quiescence (Menou et al. 1999).  Strong winds in the
thick accretion flow could also prevent much of the matter from
reaching the NS (Blandford \& Begelman 1999). Yet, it is hard to see
in both cases how this would lead to the observed asymmetric H$\alpha$
emission since the matter comes from an {\em axisymmetric} flow.\\

\section{Acknowledgments}

We are grateful to D. Mart\'\i{}nez-Delgado and G. Israelian for helpful 
comments. This work is based on observations collected at the European Southern
Observatory in Chile. This research has made use of the {\it SIMBAD} database, 
operated at CDS, Strasbourg, France. G. Dubus acknowledges support from NASA 
under grants NAG5-7007 and NAG5-7034. T. Shahbaz acknowledges support 
by the {\it Marie Curie} fellowship HP-MF-CT 199900297.

{}

\newpage

\begin{figure*}
\hspace*{2.0cm}
\psfig{file=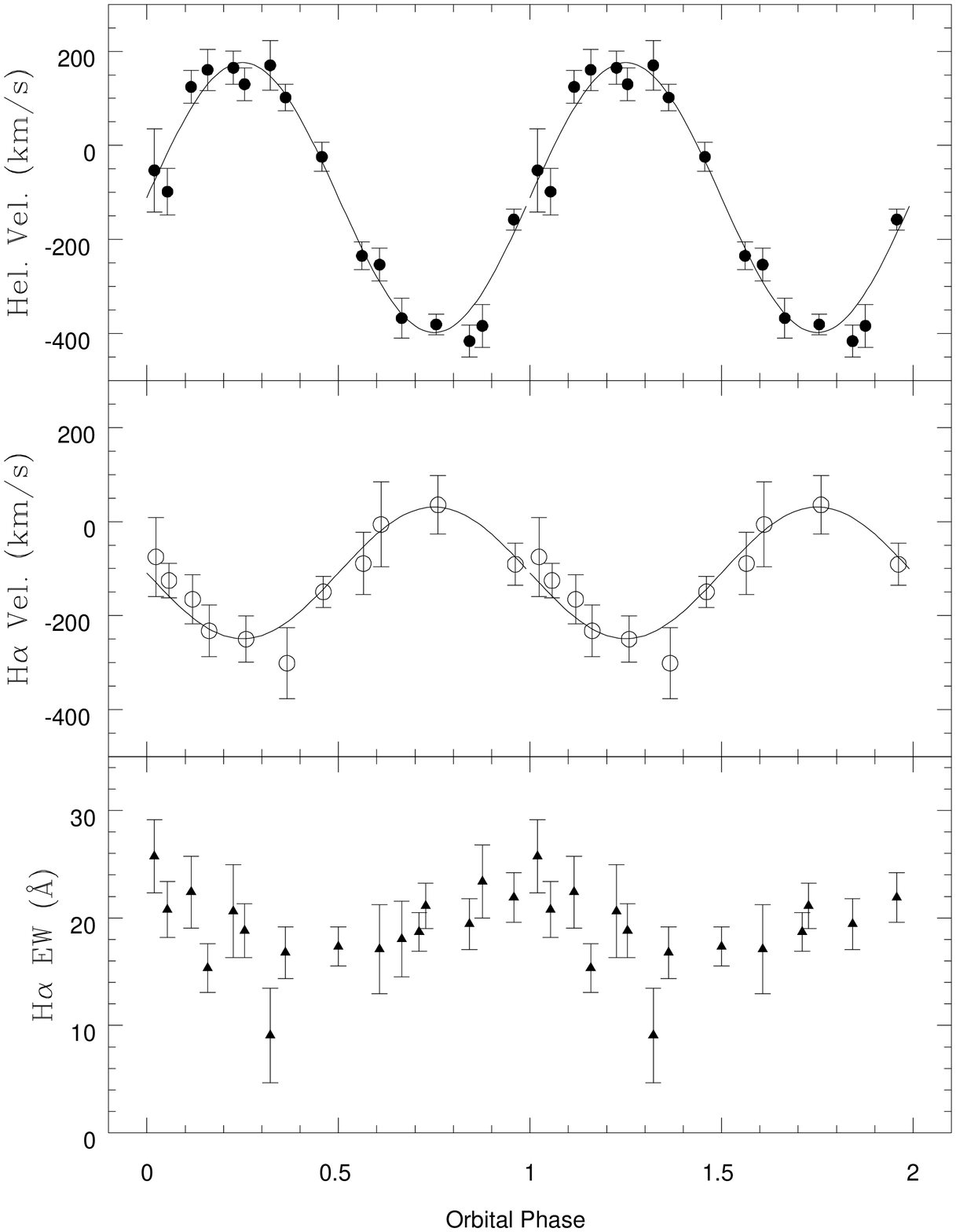,width=14cm,height=16cm,angle=0}
\caption{
Top panel: radial velocity curve of the secondary star in J2123-058. Radial
velocity points have been extracted by cross-correlating 16 phase-binned 
spectra with the K7 III template star HR 6159. 
Central panel: Radial velocity curve of the H$_{\alpha}$ emission line. 
We used a double-gaussian filter 
passband with a separation of 1400 km s$^{-1}$. 
Bottom panel: Orbital modulation of the H$_{\alpha}$ EW.}
\end{figure*}

\begin{figure*}
\hspace*{1.0cm}
\psfig{file=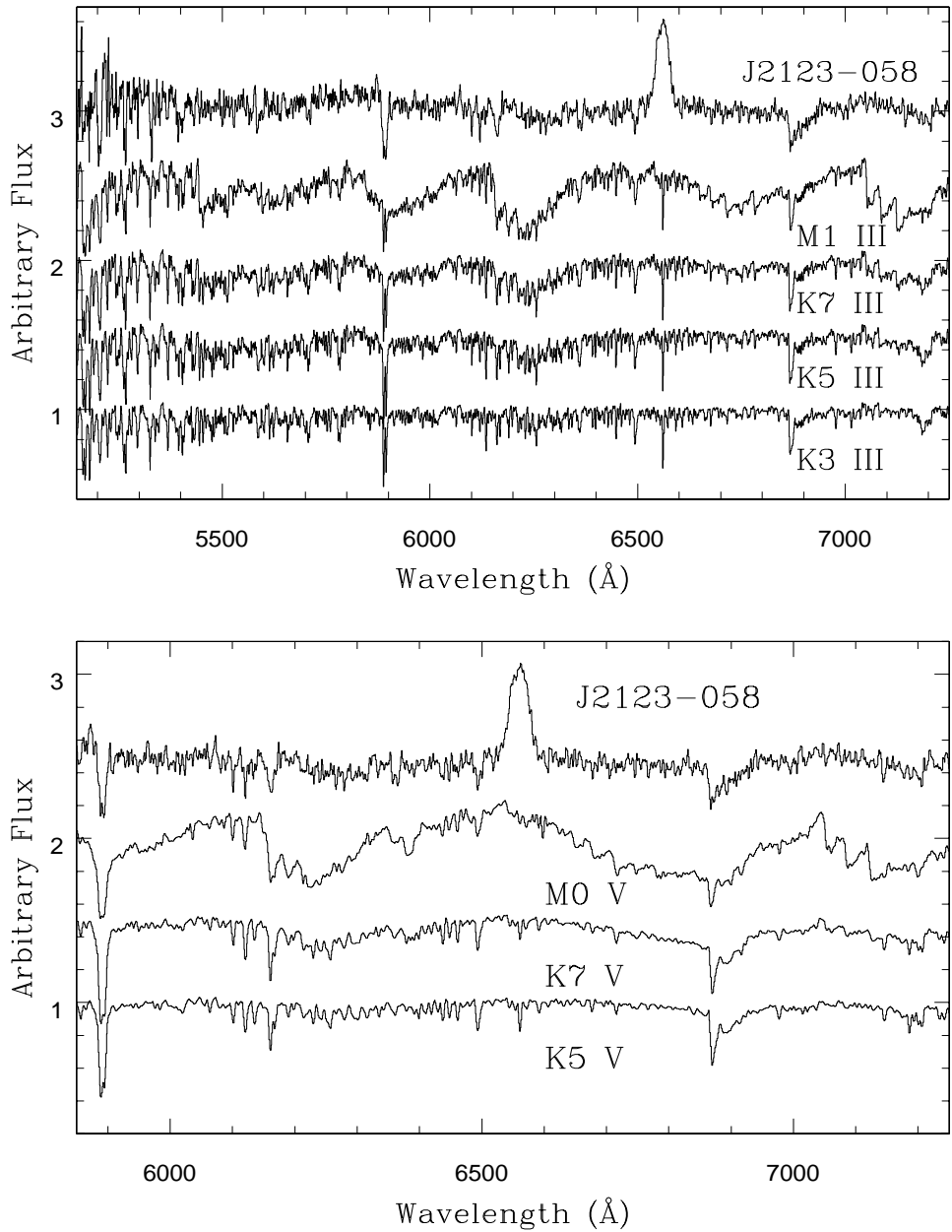,width=14cm,height=18cm,angle=0}
\caption{Doppler corrected average of J2123-058 in 
the rest frame of the secondary star together with some templates 
of luminosity classes III (upper panel) and V (lower panel). The spectrum 
of J2123-058 has been smoothed with a gaussian filter of $\sigma$=2 pixels.}
\end{figure*}

\begin{figure*}
\hspace*{1.0cm}
\psfig{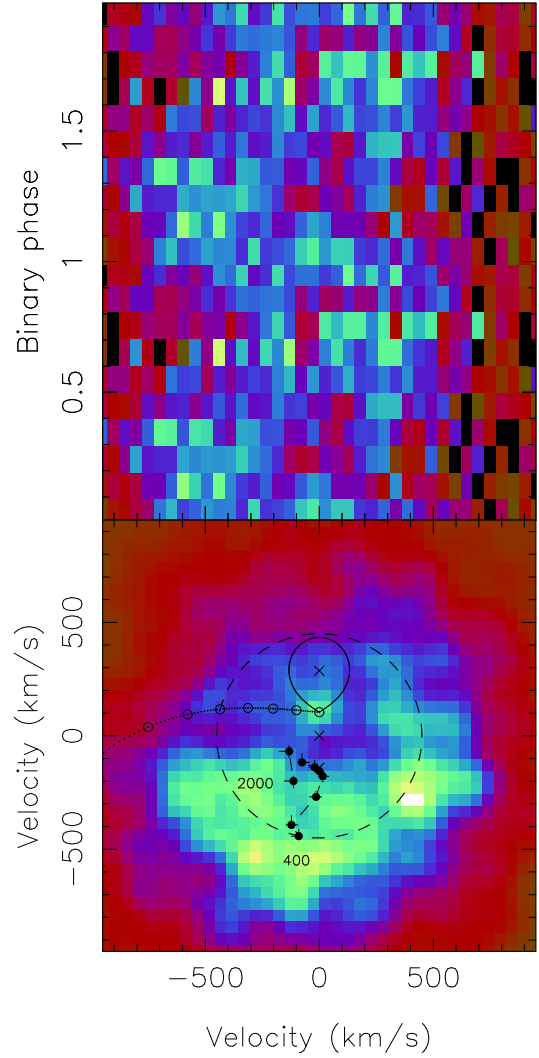}
\caption{Trailed spectra of the H$_{\alpha}$ 
profiles and the Doppler image in velocity space. The Roche lobe of the 
secondary star (for $q=0.49$) and the gas stream trajectory are indicated. 
The dashed circle marks the outer disc velocity.
We also plot the light centers obtained from the double-gaussian technique 
for a range of gaussian separations between 400 and 2000 km s$^{-1}$.}
\end{figure*}

\end{document}